\documentclass[11pt,a4paper]{article}

\usepackage{bbm}
\usepackage{verbatim}
\usepackage{amsmath}
\usepackage{amsfonts}
\usepackage{amsthm}
\usepackage{amssymb}
\usepackage{graphicx}
\usepackage[utf8]{inputenc}
\usepackage{color}
\usepackage{cite}
\usepackage{appendix}
\usepackage{hyperref}

\allowdisplaybreaks

\theoremstyle{remark}

\theoremstyle{plain}

\theoremstyle{definition}

\newcommand{\bs}[1]{{\boldsymbol{#1}}}

\begin{document}

\title{A model for the Twitter sentiment curve} 
\author{ Giacomo Aletti
\footnote{ADAMSS Center,
  Universit\`a degli Studi di Milano, Milan, Italy, giacomo.aletti@unimi.it},
Irene Crimaldi\footnote{IMT School for Advanced Studies, Lucca, Italy, 
irene.crimaldi@imtlucca.it},
Fabio Saracco
\footnote{IMT School for Advanced Studies, Lucca, Italy, 
fabio.saracco@imtlucca.it}
} 
\maketitle

\abstract{Twitter is among the most used online platforms for the political communications, due to the concision of its messages (which is particularly suitable for political slogans) and the quick diffusion of messages. Especially when the argument stimulate the emotionality of users, the content on Twitter is shared with extreme speed and thus studying the tweet sentiment if of utmost importance to predict the evolution of the discussions and the register of the relative narratives.
In this article, we present a model able to reproduce the dynamics of the sentiments of tweets 
related to specific topics and periods and to provide a prediction of the 
sentiment of the future posts based on the observed past.   
The model is a recent variant of the P\'olya urn, introduced and studied in
\cite{ale-cri-RP, ale-cri-GRP}, which is characterized by
  a ``local'' reinforcement, i.e.~a reinforcement mechanism mainly
  based on the most recent observations, and by a random persistent
  fluctuation of the predictive mean. In particular, this latter 
  feature is capable of capturing the trend fluctuations in the sentiment curve. \textcolor{black}{While the proposed model is extremely general and may be also employed in other contexts, it has been tested on several Twitter data sets and demonstrated greater performances compared to the standard P\'olya urn model. Moreover, the different performances on different data sets highlight different emotional sensitivities respect to a public event.}  
  \\[5pt]
\noindent {\em keywords:} P\'olya urn, reinforcement learning, sentiment analysis,
 urn model, Twitter.
}


\section{Introduction}
{ In the last few years, the internet has become the main source for news for citizens both in EU~\cite{TNSopinionsocial2018} and in USA~\cite{mitchell2015evolving}. Such a rapid change in the media system has created a symmetric change in the way news are delivered: before the diffusion of the web, information was \emph{intermediated} by journals, newspapers, radio and TV newscast, that represented the \emph{authority}, being publicly responsible for the diffusion of reliable news.  Nowadays, such intermediation is not present anymore: every blog or account on Facebook or Twitter assumes truthfulness just for  existing online~\cite{Zollo2015, DelVicario2016c,DelVicario2016d,Zollo2018a}. Due to this abrupt change of paradigm in the fruition of news, we observe a great increase of the diffusion of misinformation~\cite{Bradshaw2017a, GlobalDisinformation2019,NationalEndowment2019}, that appears on the web via the use of automated~\cite{Cresci2015,Ferrara2016, Shao2018a, Stella2018a, Yang2019, Cresci2019, Caldarelli2020a} or genuine accounts~\cite{DelVicario2016c, Flaxman2016,Becatti2019c,Caldarelli2020a, Pacheco2020, caldarelli2020analysis}. It has been observed that the diffusion of disinformation or misinformation campaigns leans on the emotionality of users~\cite{Zollo2015,DelVicario2016c, Qiu2017, Zollo2018a}.}\\

Twitter is
one of the most famous microblogging service, where people freely
express their views and feelings in short messages, called tweets
\cite{jansen2009}\textcolor{black}{. Twitter is reknown to be used especially for the political communications~\cite{AGCOM2017}, due to the limited amount of characters, perfectly suitable for political slogans, and for the quick sharing of messages. Due to the availability of its data, via the official API, it represents} 
an extremely rich   
resource of ``spontaneous emotional information'' \cite{ren2013}.
%
%
Sentiment analysis, also known
as opinion mining, is a collection of techniques in order to
automatically detect the positive or negative connotation of texts. 
%
%
An overview of the latest \textcolor{black}{tools,} updates and
open issues in sentiment analysis can be found in \cite{chak2020,
  patil2015, yue2019} (see also the references therein). Some examples
of applications, where predictions are formulated based on the
sentiment extracted from on-line texts are provided in \cite{bing2014,
  bollen2011, golder2011, lei2016, oconnor2010, tumasjan2010, yu2012,
  zhu2011}. In \cite{chmiel2011}, sentiment analysis is used to
investigate the emotion transmission in e-communities; while in
\cite{bollen2009}, it is employed in order to investigate on the
interplay between macroscopic socio-economic, political or cultural
events and the public mood trends, showing that these events have a
significant and immediate effect on various aspects of public mood.
The Ref. \cite{ren2013} provides a matrix-factorization method to
predict individuals' opinions toward specific topics they had not
directly given. In \cite{tan2014}, the authors consider the sentiment
curve of Twitter posts along time in order to infer the causes of
sentiment variations, leveraging on the idea that the emerging topics
discussed in the variation period could be highly related to the
reasons behind the variations. { In \cite{kursuncu2018},
the authors present the data prediction as a process based on two
different levels of granularity: i) a fine-grained analysis to make
tweet-level predictions on various aspects, such as sentiment, topics,
volume, location, timeframe, and ii) a coarse-grained analysis to
predict the outcome of a real-world event, by aggregating and
combining the fine-grained predictions. With respect to this
classification, the present work can be placed in the stream of
literature regarding the fine-grained analysis to model/predict the
sentiment of Twitter posts.
}\\ \indent  

\textcolor{black}{While an important body of research target the issue of predicting the information cascades~\cite{Chadwick2017, Zaman2014, Dow2013, Kumar2010, Kobayashi2016, Gao2015, Golosovsky2012, Zhao2015}, t}o the best of our knowledge, there are
not works that provide models for the evolution of Twitter
sentiment. We aim at filling in this gap, presenting a model that is
able to reproduce the sentiment curve of the tweets related to
specific topics and periods and to provide a prediction of the 
sentiment of the future posts based on the observed past.  
We achieve this purpose employing a recent
variant of the P\'olya urn, introduced in
\cite{ale-cri-RP} and called Rescaled P\'olya \textcolor{black}{(RP)} urn.  
{ In brief, the RP urn model differs from the standard P\'olya urn for the presence of a ``local" reinforcement, i.e. elements that are recently observed have a greater impact on the near future and may be identified as the ``fashion" of the moment. In the online social networks applications, this  local reinforcement aims at representing the persistence of an emotional response to a public event, capturing the phenomenon observed in~\cite{Zollo2015}. Moreover, it is able to correctly reproduce the sentiment dynamics of the tweets, outperforming the standard P\'olya urn model, as we will see, on several different data sets. Its  prediction ability is also quite high.  
%
%
Finally, it is worthwhile to underline that the proposed model may be also employed 
  in other contexts. \\ 
  
  \indent The sequel of the work is so structured. In
  Section \ref{sec:meth} we will present the model: after introducing the standard P\'olya model in Subsection \ref{ssec:polya}, in Subsection \ref{ssec:urn-model} we formally describe the Rescaled P\'olya urn model in general. Then, we focus on the case with two colors and, next to the general model (Complete model), we identify two special cases ("Only fashion" model and "No fashion" model).   In Section
\ref{applications}, we describe the considered datasets and we illustrate the 
performed analysis and the obtained results. Finally, in Section \ref{conclusions} we comment the results 
and draw our conclusions.
The paper is enriched by an appendix regarding the evolution of the estimated model parameters.
}

\section{Methods}\label{sec:meth}
\subsection{Standard P\'olya urn}\label{ssec:polya}
The standard P\'olya urn (see
\cite{EggPol23, mah, pemantle2007}) is a stochastic model 
driven by a reinforcement mechanism (also known as ``rich get richer" principle):
the probability that a given event occurs increases with the number of times 
the same event occurred in the past. This rule is a key feature governing the
dynamics of many biological, economic and social systems 
(see, e.g. \cite{pemantle2007}) and it seems plausible that it plays a role 
also in the sentiment dynamics of the Twitter posts as 
the emotional state of an individual influences the emotions of others 
\cite{chmiel2011, tang2012}. 
The P\'olya urn model has been widely studied and
generalized (some recent variants can be found in \cite{ale-cri-RP, ale-cri-GRP, 
  AlCrGh, ale-cri-ghi-WEIGHT-MEAN, AlGhRo, AlGhVi, BeCrPrRi16,
  cal-che-cri-pam, chen2013, collevecchio2013, Cr16, ghiglietti2017,
  lar-pag, mailler}) and in its simplest form, with $k$-colors, 
works as follows.  An urn contains $N_{0\, i}$ balls of color $i$, for
$i=1,\dots, k$, and, at each discrete time, a ball is extracted from
the urn and then it is returned inside the urn together with
$\alpha>0$ additional balls of the same color. Therefore, if we denote
by $N_{n\, i}$ the number of balls of color $i$ in the urn at time
$n$, we have
\begin{equation*}
N_{n\, i}=N_{n-1\,i}+\alpha\xi_{n\,i}=N_{0\,i}+\alpha\sum_{h=1}^n\xi_{h\,i}
\qquad\mbox{for } n\geq 1,
\end{equation*}
where $\xi_{n\,i}=1$ if the extracted ball at time $n$ is of color
$i$, and $\xi_{n\,i}=0$ otherwise. The parameter $\alpha$ regulates
the reinforcement mechanism: the greater $\alpha$, the greater the
dependence of $N_{n\,i}$ on $\sum_{h=1}^n\xi_{h\,i}$.

\subsection{Rescaled P\'olya (RP) urn}\label{ssec:urn-model}
The ``Rescaled'' P\'olya (RP) urn model, introduced in 
\cite{ale-cri-RP}, is characterized by the introduction of the
parameter $\beta$, together with the initial parameters $(b_{0\,
  i})_{i=1,\dots,k}$ and $(B_{0\,i})_{i=1,\dots,k}$, next to the
parameter $\alpha$ of the original model, so that
\begin{equation*}\label{eq-dynamics-intro}
\begin{aligned}
N_{n\, i}& =b_{0\, i}+B_{n\, i} &&\text{with }
\\
B_{n\, i}&=\beta B_{n-1\, i}+\alpha\xi_{n\, i}&& n\geq 1.
\end{aligned}
\end{equation*}
Therefore, the urn initially contains $b_{0\,i}+B_{0\,i}>0$ balls of
color $i$ and the parameter $\beta\geq 0$, together with $\alpha>0$,
regulates the reinforcement mechanism. More precisely, the term $\beta
B_{n-1\,i}$ links $N_{n\,i}$ to the ``configuration'' at time $n-1$
through the ``scaling'' parameter $\beta$, and the term
$\alpha\xi_{n\,i}$ links $N_{n\,i}$ to the outcome of the extraction
at time $n$ through the parameter $\alpha$.  Note that the case
$\beta=1$ corresponds to the standard P\'olya urn with an initial
number $N_{0\,i}=b_{0\,i}+B_{0\,i}$ of balls of color $i$.  When
$\beta\in [0,1)$, this variant of the P\'olya urn is characterized by
  a ``local'' reinforcement, i.e.~a reinforcement mechanism mainly
  based on the most recent observations, and by a random persistent
  fluctuation of the predictive mean
  $\psi_{n\,i}=E[\xi_{n+1\,i}=1|\,\mbox{``past''}]$. As we will show,
  this latter feature is capable of capturing the trend fluctuations in
  the sentiment curve  of Twitter posts 
  (see Figs. \ref{migration-all-fitted}-\ref{covid-onlyBOT-fitted}). 
  \\
  
  
\indent More formally, given a vector $\bs{x}= (x_1,\ldots, x_k)^\top\in \mathbb{R}^k$, 
we define $|\bs{x}| = \sum_{i=1}^k
|x_i|$. Moreover, we set
$\bs{b_0}=(b_{0\,1},\dots,b_{0\,k})^{\top}$ and
$\bs{B_0}=(B_{0\,1},\dots,B_{0\,k})^{\top}$, we assume $|\bs{b_0}|>0$ and we define $\bs{p_0} =
\frac{\bs{b_0}}{|\bs{b_0}|}$.  At each discrete time $(n+1)\geq 1$, a
ball is drawn at random from the urn, obtaining the random vector
$\bs{\xi_{n+1}} = (\xi_{n+1\,1}, \ldots, \xi_{n+1\,k})^\top$ defined
as
\begin{equation*}
\xi_{n+1\,i} = 
\begin{cases}
1  &  \text{when the extracted ball at time $n+1$ is of color $i$}
\\
0  & \text{otherwise}\,.
\end{cases}
\end{equation*}
The number of balls in the urn is so updated:
\begin{equation}\label{eq:reinf1:K}
\bs{N_{n+1}}=\bs{b_0}+\bs{B_{n+1}}\qquad\text{with}
\qquad
\bs{B_{n+1}} = \beta \bs{B_n} + \alpha \bs{\xi_{n+1}}\,,
\end{equation}
which gives (since $|\bs{\xi_{n+1}}|=1$)
\begin{equation}\label{eq:reinf1:K-bis}
|\bs{B_{n+1}}|= \beta |\bs{B_n}| + \alpha.  
\end{equation}
Therefore, setting $r^*_{n} = |\bs{N_n}|= |\bs{b_0}|+|\bs{B_n}|$, we
get
\begin{equation}\label{dinamica-rnstar}
r_{n+1}^*=r_n^*+(\beta-1)|\bs{B_n}|+\alpha.
\end{equation}
Moreover, denoting by $\mathcal{F}=(\mathcal{F}_n)_{n\geq 0}$ the
filtration representing the information along time (formally, this
means to set $\mathcal{F}_0$ equal to the trivial $\sigma$-field and
$\mathcal{F}_n=\sigma(\bs{\xi_1},\dots,\bs{\xi_n})$ for $n\geq 1$), the
conditional probabilities $\bs{\psi_{n}}= (\psi_{n\,1}, \ldots,
\psi_{n\,k})^\top$ of the extraction process, also called predictive
means, are
\begin{equation}\label{eq:extract1a:K}
\psi_{n\,i} =E[\xi_{n+1\,i}\,|\,\mathcal{F}_n]=P( \xi_{n+1\,i} = 1| \mathcal{F}_n)= 
\frac{N_{n\,i}}{|\bs{N_n}|}=
\frac{b_{0\, i}+B_{n\, i}}{r_n^*}, \qquad i=1, \ldots k,\; n\geq 0\,.
\end{equation}
This urn model has been studied in \cite{ale-cri-RP, ale-cri-GRP}. 
All the mathematical proofs and details can be found in these papers.

\subsubsection{Two colors ($k=2$)}
With two colors, the quantity of interest are only
$\xi_{n}=\xi_{n\,1}=1-\xi_{n\,2}$ and $\psi_{n} = \psi_{n\,1} =
1-\psi_{n\,2}$.  In the sequel, we consider the RP urn model with $\beta=1$
(i.e. the standard P\'olya urn model) and with $\beta<1$. In the first case, we have 
$$
\psi_{n}=\frac{N_{0\,1}+\alpha\sum_{h=1}^n \xi_{h}}{|\bs{N_0}|+\alpha n}\,.
$$
In the second case, by
\eqref{eq:reinf1:K}, \eqref{eq:reinf1:K-bis}, \eqref{dinamica-rnstar}
and \eqref{eq:extract1a:K}, using $\sum_{m=0}^{n-1}
x^m=(1-x^{n})/(1-x)$, we obtain
\begin{equation*}\label{eq-rstar_n}
r_n^*=|\bs{b_0}|+\frac{\alpha }{1-\beta}+
\beta^n\left(|\bs{B_0}|- \frac{\alpha }{1-\beta}\right)
\longrightarrow r^*=|\bs{b_0}|+\frac{\alpha}{1-\beta}
\end{equation*}
and 
\begin{equation*}\label{eq-psi_n}
\psi_n = 
\frac{ 
b_{0\, 1}+ \beta^n B_{0\,1} + \alpha \sum_{h=1}^n \beta^{n-h} \xi_{h}}
{
|\bs{b_0}|+\frac{\alpha }{1-\beta}+
\beta^n\big(|\bs{B_0}|- \frac{\alpha }{1-\beta}\big)}\,.
\end{equation*}
Since $\beta<1$, the dependence of $\psi_n$ on $\xi_h$ exponentially
increases with $h$, because of the factor $\beta^{n-h}$, and so the
main contribution is given by the most recent extractions. We refer to
this phenomenon as ``local'' reinforcement. The case $\beta=0$ is an
extreme case, for which $\psi_n$ depends only on the last extraction
$\xi_n$. Note that, when $\beta=1$, i.e. the case of the standard P\'olya urn, 
all the past observations $\xi_h$ equally contribute to $\psi_n$, 
with a weight equal to $\alpha$. This different dependence on the past 
leads to a different behaviour of $\psi_n$ along time (see \cite{ale-cri-RP}): 
in the standard P\'olya urn, the process $(\psi_n)$  asymptotically stabilizes, 
converging almost surely toward a random variable, while in the RP urn, the process
$(\psi_n)$ persistently fluctuates 
(see Figs. \ref{migration-all-fitted}-\ref{covid-onlyBOT-fitted}).  
\\ \indent If we set 
\begin{equation*}
p_0 =p_{0\,1}=\frac{b_{0\, 1}}{|\bs{b_0}|}, \quad 
(1-\gamma^*) = \frac{|\bs{b_0}|}{ r^*},
\quad \widetilde{B}_n = \frac{{B}_{n\,1}}{|\bs{B_n}|}\,,
\end{equation*}
we get for a large $n$
\begin{equation*}
  \begin{split}
\psi_{n+1}&=\frac{b_{0\,1}}{r_{n+1}^*}+\frac{B_{n+1\,1}}{r_{n+1}^*}=
\frac{|\bs{b_0}|}{r_{n+1}^*}p_0+\frac{|\bs{B_{n+1}}|}{r_{n+1}^*}\widetilde{B}_{n+1}
\\
&=
\frac{|\bs{b_0}|}{r_{n+1}^*}p_0+\frac{r_{n+1}^*-|\bs{b_0}|}{r_{n+1}^*}
\widetilde{B}_{n+1}
\\
&\sim \frac{|\bs{b_0}|}{r^*}p_0+\frac{r^*-|\bs{b_0}|}{r^*}\widetilde{B}_{n+1}
=(1-\gamma^*) p_0  + \gamma^* \widetilde{B}_{n+1}
\end{split}
\end{equation*}
and
\begin{equation*}
  \begin{split}
    \widetilde{B}_{n+1}&=\frac{B_{n+1}}{|\bs{B_{n+1}}|}=
    \frac{\beta}{|\bs{B_{n+1}}|}
    B_n+\frac{\alpha}{|\bs{B_{n+1}}|}\xi_{n+1}\\
    &=\beta\frac{r_n^*-|\bs{b_0}|}{r_{n+1}^*-|\bs{b_0}|}\widetilde{B}_n
    +\frac{\alpha}{r_{n+1}^*-|\bs{b_0}|}\xi_{n+1}\\
    &\sim \beta\widetilde{B}_n+\frac{\alpha}{r^*-|\bs{b_0}|}\xi_{n+1}
    =\beta\widetilde{B}_n+(1-\beta)\xi_{n+1}\,.
    \end{split}
\end{equation*}
Summing up, the model dynamics can be approximated for $n$ large by 
\begin{equation*}
\psi_{n+1} = (1-\gamma^*) p_0  + \gamma^* \widetilde{B}_{n+1} , \qquad 
\widetilde{B}_{n+1} = \beta \widetilde{B}_{n} + (1-\beta) \xi_{n+1},
\end{equation*}
where $p_0 ,\, \gamma^*,\, \beta,\, \widetilde{B}_{0}$ are
the parameters. Note that $\alpha$ does not appear among the
parameters for the above approximated dynamics, but it is included in the new parameter $\gamma^*$. 
Moreover, the quantity
$\widetilde{B}_0$ is exponentially fast negligible, because we have
$\widetilde{B}_n=\beta^n\widetilde{B}_0+(1-\beta)\sum_{k=1}^n\beta^{n-k}\xi_k$, with $\beta<1$.
Therefore, the fundamental parameters are $p_0,\, \gamma^*$ and
$\beta$: $p_0$ is a deterministic component, $\gamma^*$ tunes the
weight in the predictive mean $\psi_{n+1}$ of the random
``fluctuation'' component $\widetilde{B}_{n+1}$ with respect to the
deterministic one, and $\beta$ regulates the dependence of the present state
$\widetilde{B}_{n+1}$ on the previous state $\widetilde{B}_n$ and on
the present observation $\xi_{n+1}$. We refer to $(\widetilde{B}_n)_n$ as the "fashion" process, since it 
reproduces the trend variations of the considered phenomenon (in our case, the sentiment of Twitter posts) . 
In the following applications, we consider the following cases:
\begin{itemize}
\item {\em Complete RP model:} The three parameters
  $\theta=(p_0,\gamma^*,\beta)$ are free to vary.
\item{\em ``Only Fashion'' RP model:} $\gamma^* = 1$ (and $p_0 = 0$
  irrelevant). This means that the predictive mean is not driven by
  any deterministic component, but it coincides with the fashion process. 
  The free parameter is given by $\theta=\beta$.
\item{\em "No Fashion" RP model:} $\gamma^* = 0$ (and
  $\beta = 0$ irrelevant). In this case $\psi_n$ is equal to the constant
  $p_0$ and, consequently, the free parameter is given by
  $\theta=p_0$. 
\end{itemize}

\section{Results}\label{applications}

\subsection{Data}\label{data}
Data have been collected from the Twitter platform, using the official API to Stream the exchange of messages on several topics. In the following, the various datasets are described in more details. 
\\

\begin{itemize}
\item {{\bf Italy, Migration debate}}\\
Data were collected through the Filter API since 23rd of January to 22nd of February 2019 and targeted the Italian debate on migration. 
Data were previously analysed in~\cite{Caldarelli2020a}.
In the dataset, the information about the nature, automated or not, of the users is present. The bot detection algorithm embedded is a lightweight version of the classifier proposed in~\cite{Cresci2015}; more details on the dataset can be found in~\cite{Caldarelli2020a}.

\item {\bf Italy, 10 days of traffic}\\
The dataset collects the entire traffic, compatibly with the Filter API sampling, of messages in Italian in the days from the first to the 10th of September 2019: the keyword used for the query were the Italian vowels, in order to collect all messages that may contain some word. The bot detection algorithm used was developed in~\cite{Yang2019}.

\item{\bf Italy, COVID-19 epidemic}\\ 
The dataset covers the period from February 21st to April to 20th 2020, including tweets in Italian language, and was previously analysed in~\cite{caldarelli2020analysis}. The keywords used for the query are relative to the COVID-19 epidemic; more details can be found in the original reference. The dataset includes information on the automated or not nature of the accounts, detected using the algorithm developed in~\cite{Yang2019}. 
\end{itemize}

For every message, the relative sentiment was calculated using the \emph{polyglot} python module developed in~\cite{chen2014building}. This module  provides a numerical value $v$ for the sentiment and we fix a threshold $T=0.35$ so that we classify as a
tweet with positive sentiment those with $v>T$ and as a tweet with
negative sentiment those with $v <-T$. We discard tweets with a value $v\in [-T,T]$. 
Tables \ref{migration-descr}-\ref{covid-descr} show some descriptives of the considered datasets:\\

\begin{table}[htb]
\begin{footnotesize}
  \begin{center}
\caption{``Migration" dataset: Descriptives.}
\label{migration-descr}
\begin{tabular}{|r|c|c|c||}
\hline
\textbf{Migration} &\textbf{Entire}  & \textbf{Only BOTs' posts}  
\\
\hline
Posts & 367367 & 4124 
\\
Percentage of positive posts & 49.60\% & 47.97\% 
\\ 
\hline
\end{tabular}
\end{center}
\end{footnotesize}
\end{table}

\begin{table}[htb]
\begin{footnotesize}
  \begin{center}
\caption{``10 days of traffic" dataset: Descriptives.}
\label{traffic-descr}
\begin{tabular}{|r|c|c|c|}
\hline
\textbf{10 days of traffic} &\textbf{Entire}  & \textbf{Only BOTs' posts}  
\\
\hline
Posts &  3164620 &  102374
\\
Percentage of positive posts & 63.26\% &  63.79\%
\\ 
\hline
\end{tabular}
\end{center}
\end{footnotesize}
\end{table}

\begin{table}[htbp]
\begin{footnotesize}
  \begin{center}
\caption{``Covid" dataset: Descriptives.}
\label{covid-descr}
\begin{tabular}{|r|c|c|c|}
\hline
\textbf{Covid} &\textbf{Entire}  & \textbf{Only BOTs' posts}  
\\
\hline
Posts &  2037584 &  48252
\\
Percentage of positive posts & 50.58\% &  54.00\%
\\ 
\hline
\end{tabular}
\end{center}
\end{footnotesize}
\end{table}

\subsection{Analysis of the prediction ability}\label{method}
We apply the RP model with $k=2$: the time series of the tweets
represents the time series of the extractions from the urn, that is
the random variables $\xi_{n}$. The event $\{\xi_n=1\}$ means that
tweet $n$ exhibits a positive sentiment, while $\{\xi_n=0\}$ means
that tweet $n$ exhibits a negative sentiment.  The parameters have
been estimated by maximum likelihood. More precisely, we have divided
the observations into $S$ slots.  For each slot $s = 1,\ldots,S-1$,
with training data of the slots $t= 0,\ldots,s-1$, we have estimated
the best parameters $\widehat{\theta}(s)$ for the different proposed 
models: Standard P\'olya, Complete RP, Only Fashion RP, No Fashion RP. (See Appendix, Sec. \ref{appendix-parameters} for a further discussion.) 
With these parameters, for each observation $\xi_{n+1}$ of the $s$-th
slot, we have estimated the probability $\widehat{\psi}_n$, that is the predictive mean, 
as a function of the estimated parameters and of the observed history
$\xi_1, \ldots,\xi_{n}$.  The predictive mean $\widehat{\psi}_n$ represents  
our prediction of the future value $\xi_{n+1}$.  We have quantified the ability 
of the considered model to predict the future outcomes 
by means of the relative squared error  
with respect to the method that predicts the future outcome 
taking the value assumed by the majority in the past. 
More precisely, we have computed the following quantity:
\begin{equation}\label{confronto}
SS_{rel}=\frac{\sum_{s=1}^{S-1}\sum_{n=s[N/S]}^{(s+1)[N/S]-1} (\xi_{n+1}-m_s)^2}
{\sum_{s=1}^{S-1} \sum_{n=s [N/S]}^{(s+1)[N/S]-1} (\xi_{n+1}-\widehat{\psi}_n)^2}
\,,
\end{equation}
where $N$ is the size of the dataset and 
$m_s$ is the value assumed by the majority in the slots $0,\dots, s-1$.
This quantity measures the ability of the model to predict the future
outcomes: the greater it is, the better is the prediction with respect to the 
method based on the past majority. The values $SS_{rel}$ obtained for the different 
considered models are also compared with the ``theoretical" value 
of $SS_{rel}$ computed replacing $\widehat{\psi}_n$ by 
$\bar{\psi} = \sum_{n=[N/S]+1}^N\xi_n/(N-[N/S])$, that is 
the mean value on all the period between $[N/S]+1$ and $N$. 
The term ``theoretical" is used in order to point out that $\bar{\psi}$ is of course  
not a prediction, but it gives the \emph{a posteriori} best constant fit 
once we have collected all the data until time $N$.  Summing up, our aim is 
twofold: to obtain a value of $SS_{rel}$ greater than $1$, that means that 
the considered models beat the performance of the method based on the majority, and to 
get a value greater or equal to the ``theoretical" value, that means that 
the proposed models perform better or similarly than the (theoretical) 
\emph{a posteriori} best constant fit.
\\
\indent We summerize the results in Tables \ref{migration-analisi-comp}-\ref{covid-analisi-comp}. 
For each considered dataset, 
we have also analysed the subset obtained
by the restriction to the tweets classified as sent by a bot. 
In the tables the best values are highlighted in bold. We can observe that,  
for the ``Migration" and ``Covid" datasets, the considered models perform 
more or less two times better than the method based on the majority and 
this performance is similar to  (indeed, in the most cases slightly better than) 
the one given by the (theoretical) \emph{ a posteriori} best constant fit.
For the ``10 days traffic" dataset, the performance  of the considered models  
is one and half times better  than the method based on the majority and 
this performance is similar to the one given by the (theoretical)  
\emph{a posteriori} best constant fit. Moreover, the performances of the 
``Complete RP" model and of the ``Only fashion RP" model do not significantly differ; while 
the "No Fashion RP" model performs less well. Therefore, in the next subsection, we will discard 
this last model. 
\\

\begin{table}[htb]
\begin{footnotesize}
  \begin{center}
\caption{``Migration" dataset: Comparison of the different considered models in terms of
  \eqref{confronto}.}
\label{migration-analisi-comp}
\begin{tabular}{|r|c|c|c|c|c|}
\hline
\textbf{Migration} &\textbf{Standard P\'olya}  & \textbf{Complete RP}  
& \textbf{Only Fashion RP} & \textbf{No Fashion RP} &\textbf{Theoretical value}
\\
\hline
Entire & 199.96\% & {\bf 205.30\%} & 205.25\% & 199.83\% & 199.98\% 
\\
Only BOTs & 192.55\% & 197.62\% & {\bf 197.74\%} & 192.28\% & 192.77\% 
\\ 
\hline
\end{tabular}
\end{center}
\end{footnotesize}
\end{table}

\begin{table}[htb]
\begin{footnotesize}
  \begin{center}
\caption{``10 days of traffic" dataset: Comparison of the different considered models in terms of
  \eqref{confronto}.}
\label{traffic-analisi-comp}
\begin{tabular}{|r|c|c|c|c|c|}
\hline
\textbf{10 days of traffic} &\textbf{Standard P\'olya}  & \textbf{Complete RP}  
& \textbf{Only Fashion RP} & \textbf{No Fashion RP} &\textbf{Theoretical value}
\\
\hline
Entire & 159.29\% & 159.43\% & {\bf 159.43\%} & 159.29\% & 159.29\% 
\\
Only BOTs & 157.99\% & {\bf 158.12\%} & 158.04\% & 157.99\% & 158.00\% 
\\
\hline
\end{tabular}
\end{center}
\end{footnotesize}
\end{table}

\begin{table}[htb]
\begin{footnotesize}
  \begin{center}
\caption{``Covid" dataset: Comparison of the different considered models
  in terms of \eqref{confronto}.}
\label{covid-analisi-comp}
\begin{tabular}{|r|c|c|c|c|c|}
\hline
\textbf{Covid} &\textbf{Standard P\'olya}  & \textbf{Complete RP}  
& \textbf{Only Fashion RP} & \textbf{No Fashion RP} &\textbf{Theoretical value}
\\
\hline 
Entire & 198.41\% & {\bf 201.57\%} & 201.56\% & 198.40\% & 198.46\% 
\\
Only BOTs & 186.10\% & 188.91\% & {\bf 188.94\%} & 186.09\% & 186.35\% 
\\
\hline
\end{tabular}
\end{center}
\end{footnotesize}
\end{table}


\subsection{Fluctuations of the sentiment curve}
\label{subsec:fluctuations}
\begin{figure}[htb]
\begin{center}
	\includegraphics[width=0.55\textwidth]{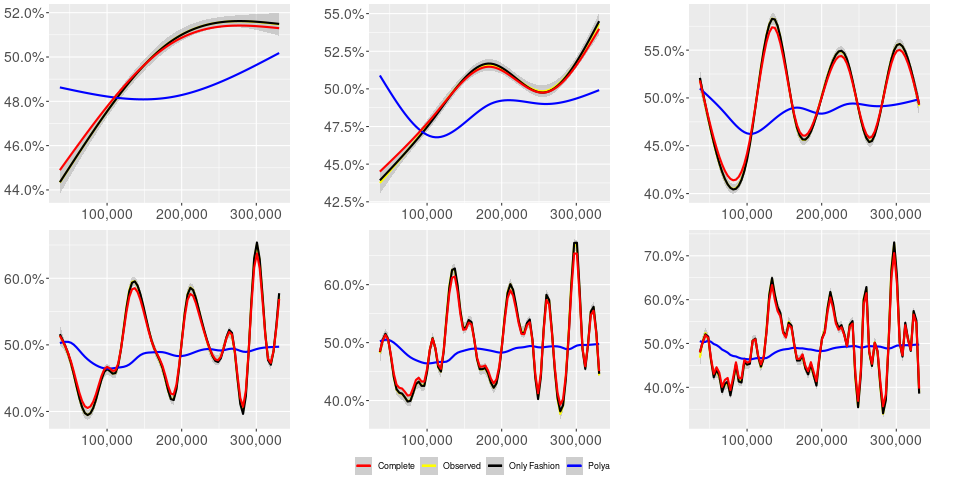}
\end{center}
	\caption{``Migration" dataset (Entire): Sentiment curves}
\label{migration-all-fitted}
\end{figure}

\begin{figure}[htb]
\begin{center}
	\includegraphics[width=0.55\textwidth]{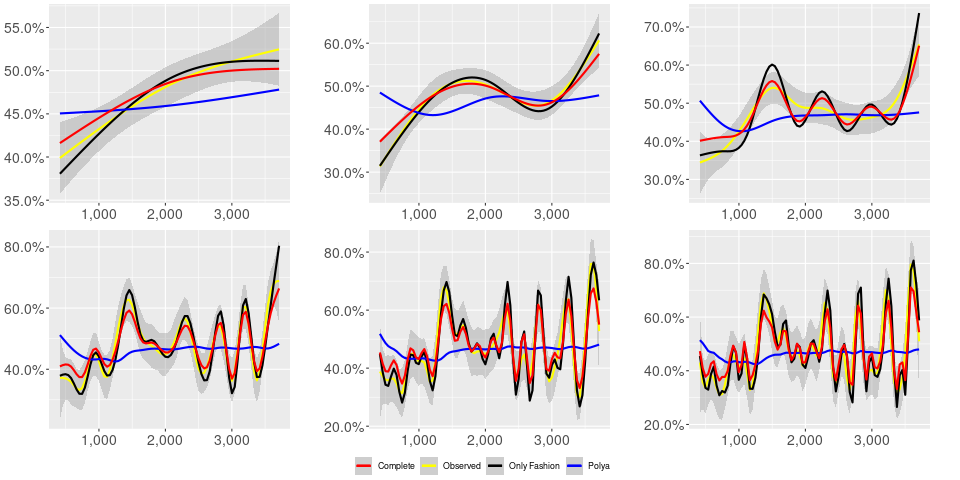}
\end{center}
	\caption{``Migration", automated account subset: Sentiment curves for BOTs' posts}
\label{migration-onlyBOT-fitted}
\end{figure}

\begin{figure}[htb]
\begin{center}
	\includegraphics[width=0.55\textwidth]{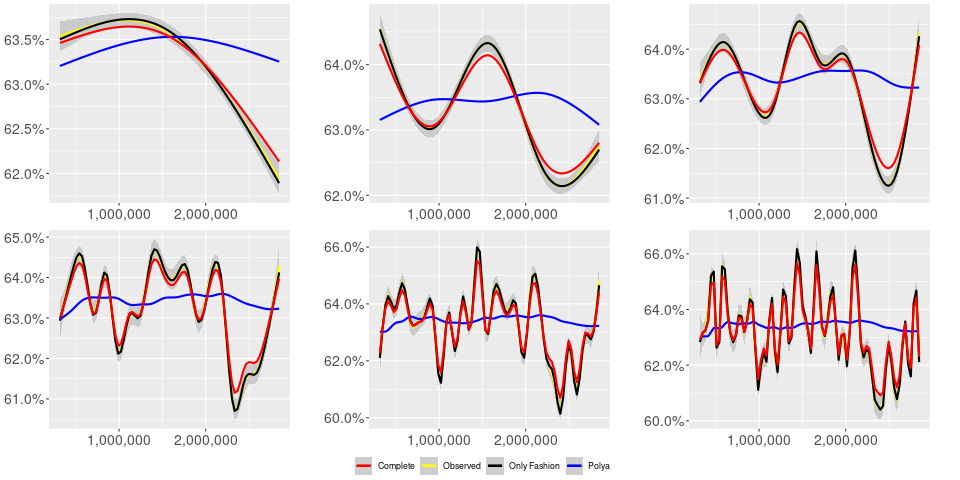}
\end{center}
\caption{``10 days of traffic" dataset (Entire): Sentiment curves }
\label{traffic-all-fitted}
\end{figure}

\begin{figure}[htb]
\begin{center}
	\includegraphics[width=0.55\textwidth]{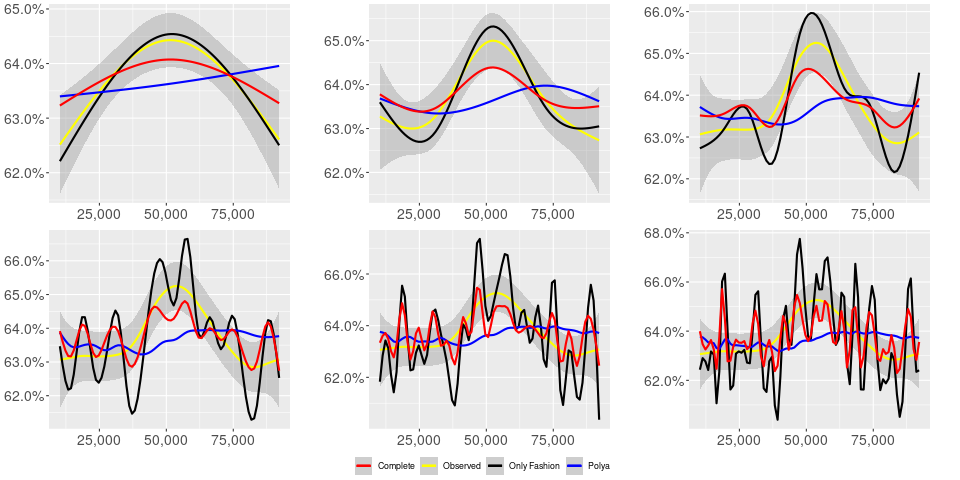}
\end{center}
	\caption{``10 days of traffic", automated account subset: Sentiment curves for BOTs' posts}
\label{traffic-onlyBOT-fitted}
\end{figure}

\begin{figure}[htb]
\begin{center}
\includegraphics[width=0.55\textwidth]{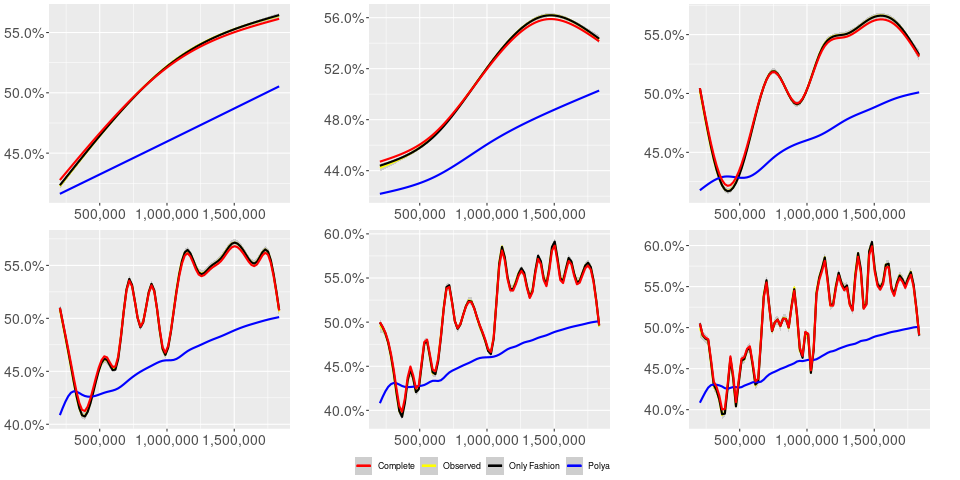}
\end{center}
\caption{``Covid" dataset (Entire): Sentiment curves}
\label{covid-all-fitted}
\end{figure}

\begin{figure}[htb]
\begin{center}
\includegraphics[width=0.55\textwidth]{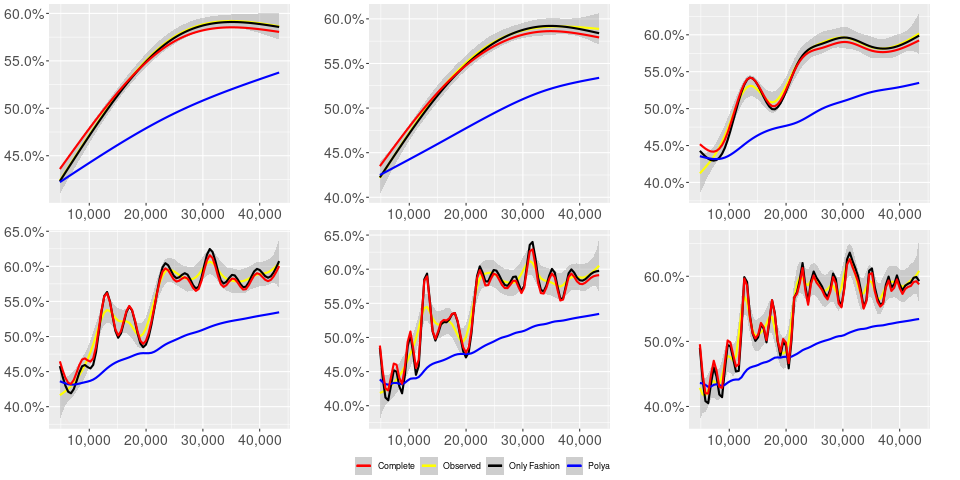}
\end{center}
	\caption{``Covid", automated account subset: Sentiment curves for BOTs' posts}
\label{covid-onlyBOT-fitted}
\end{figure}

We provide some tables and figures in order to point out how the different considered models are able
to reproduce the trend fluctuation of the sentiment curve. More
precisely, in Figures \ref{migration-all-fitted}-\ref{covid-onlyBOT-fitted}, the yellow line is the cubic
spline smoothing (with different numbers of nodes$= 3,\,5,\,10,\,20,\,30,\,50$) 
of the time series of the observed tweets $\xi_{n+1}$,
together with the default confidence interval (gray), the red line
represents the cubic spline smoothing (with the same number of nodes) 
of the time series of the estimated predictive means 
$\widehat{\psi}_n$ (see Subsec. \ref{method}), obtained with the complete RP model, the black and
the blue lines provide similar approximations obtained with the other
models: black=Only fashion RP model and blue=Standard P\'olya model. In Tables \ref{migration-all-smoothing-table}-\ref{covid-onlyBOT-smoothing-table}, we compare the different models by means of the mean squared error (MSE), i.e.
\begin{equation}
    MSE=\frac{\sum_{n=[N/S]}^{S[N/S]-1} (\xi_{n+1}-\widehat{\psi}_n)^2}{(S-1)[N/S]}\,,
\end{equation}
where $\xi_{n+1}$ and $\widehat{\psi}_n$ refer to the values on the curves with a given smoothing. 
\\
\indent We can observe that, as explained before in Section \ref{ssec:urn-model}, the RP urn model is able to 
reproduce the fluctuations of the observed sentiment curve, while the 
standard P\'olya urn model produces a curve that converges to a value. 

\begin{table}[htb]
\begin{footnotesize}
  \begin{center}
\caption{"Migration" dataset (Entire): MSE for different levels of smoothing}
\label{migration-all-smoothing-table}
\begin{tabular}{lrrr}
  \hline
smooth & Only Fashion RP & Complete RP & Standard P\'olya \\ 
  \hline
no smooth & $2.44 \times 10^{-1}$ & $\mathbf{2.43 \times 10^{-1}}$ & $2.50 \times 10^{-1}$ \\ 
  k = 3 & $\mathbf{3.44 \times 10^{-9}}$ & $1.41 \times 10^{-6}$ & $3.03 \times 10^{-4}$ \\ 
  k = 5 & $\mathbf{1.19 \times 10^{-8}}$ & $3.23 \times 10^{-6}$ & $3.43 \times 10^{-4}$ \\ 
  k = 10 & $\mathbf{2.64 \times 10^{-7}}$ & $1.74 \times 10^{-5}$ & $1.64 \times 10^{-3}$ \\ 
  k = 20 & $\mathbf{1.04 \times 10^{-6}}$ & $2.98 \times 10^{-5}$ & $2.73 \times 10^{-3}$ \\ 
  k = 30 & $\mathbf{2.79 \times 10^{-6}}$ & $4.03 \times 10^{-5}$ & $3.83 \times 10^{-3}$ \\ 
  k = 50 & $\mathbf{7.18 \times 10^{-6}}$ & $5.41 \times 10^{-5}$ & $4.85 \times 10^{-3}$ \\ 
   \hline
\end{tabular}
\end{center}
\end{footnotesize}
\end{table}

\begin{table}[htb]
\begin{footnotesize}
  \begin{center}
\caption{"Migration" dataset (only Bots' posts): MSE for different levels of smoothing}
\label{migration-onlyBOT-smoothing-table}
\begin{tabular}{lrrr}
  \hline
smooth & Only Fashion RP & Complete RP & Standard P\'olya \\ 
  \hline
no smooth & $2.43 \times 10^{-1}$ & $\mathbf{2.41 \times 10^{-1}}$ & $2.50 \times 10^{-1}$ \\ 
  k = 3 & $\mathbf{2.62 \times 10^{-6}}$ & $3.56 \times 10^{-5}$ & $7.66 \times 10^{-4}$ \\ 
  k = 5 & $4.19 \times 10^{-4}$ & $\mathbf{1.90 \times 10^{-4}}$ & $1.10 \times 10^{-3}$ \\ 
  k = 10 & $\mathbf{1.03 \times 10^{-4}}$ & $3.50 \times 10^{-4}$ & $3.36 \times 10^{-3}$ \\ 
  k = 20 & $\mathbf{5.36 \times 10^{-4}}$ & $8.19 \times 10^{-4}$ & $6.58 \times 10^{-3}$ \\ 
  k = 30 & $\mathbf{9.09 \times 10^{-4}}$ & $1.22 \times 10^{-3}$ & $9.13 \times 10^{-3}$ \\ 
  k = 50 & $2.53 \times 10^{-3}$ & $\mathbf{2.36 \times 10^{-3}}$ & $1.34 \times 10^{-2}$ \\ 
   \hline
\end{tabular}
\end{center}
\end{footnotesize}
\end{table}

\begin{table}[htb]
\begin{footnotesize}
\begin{center}
\caption{``10 days of traffic" dataset (Entire): MSE for different levels of smoothing}
\label{traffic-all-smoothing-table}
\begin{tabular}{lrrr}
  \hline
smooth & Only Fashion RP & Complete RP & Standard P\'olya \\ 
  \hline
no smooth & $2.32 \times 10^{-1}$ & $2.32 \times 10^{-1}$ & $2.33 \times 10^{-1}$ \\ 
  k = 3 & $\mathbf{3.15 \times 10^{-9}}$ & $2.61 \times 10^{-7}$ & $1.22 \times 10^{-5}$ \\ 
  k = 5 & $\mathbf{3.86 \times 10^{-9}}$ & $8.09 \times 10^{-7}$ & $3.34 \times 10^{-5}$ \\ 
  k = 10 & $\mathbf{1.94 \times 10^{-8}}$ & $2.02 \times 10^{-6}$ & $6.88 \times 10^{-5}$ \\ 
  k = 20 & $\mathbf{7.81 \times 10^{-8}}$ & $2.65 \times 10^{-6}$ & $8.80 \times 10^{-5}$ \\ 
  k = 30 & $\mathbf{1.74 \times 10^{-7}}$ & $2.86 \times 10^{-6}$ & $9.65 \times 10^{-5}$ \\ 
  k = 50 & $\mathbf{1.08 \times 10^{-6}}$ & $5.15 \times 10^{-6}$ & $1.53 \times 10^{-4}$ \\ 
   \hline
\end{tabular}
\end{center}
\end{footnotesize}
\end{table}

\begin{table}[htb]
\begin{footnotesize}
\begin{center}
\caption{"10 days traffic" dataset (only BOTs' posts): MSE for different levels of smoothing}
\label{traffic-onlyBOT-smoothing-table}
\begin{tabular}{lrrr}
  \hline
smooth & Only Fashion RP & Complete RP & Standard P\'olya \\ 
  \hline
no smooth & $2.31 \times 10^{-1}$ & $2.31 \times 10^{-1}$ & $2.31 \times 10^{-1}$ \\ 
  k = 3 & $\mathbf{4.10 \times 10^{-7}}$ & $6.67 \times 10^{-7}$ & $5.73 \times 10^{-6}$ \\ 
  k = 5 & $\mathbf{7.95 \times 10^{-7}}$ & $2.02 \times 10^{-5}$ & $5.97 \times 10^{-5}$ \\ 
  k = 10 & $\mathbf{6.81 \times 10^{-6}}$ & $2.35 \times 10^{-5}$ & $7.19 \times 10^{-5}$ \\ 
  k = 20 & $\mathbf{1.59 \times 10^{-5}}$ & $5.43 \times 10^{-5}$ & $1.52 \times 10^{-4}$ \\ 
  k = 30 & $\mathbf{2.59 \times 10^{-5}}$ & $5.98 \times 10^{-5}$ & $1.75 \times 10^{-4}$ \\ 
  k = 50 & $\mathbf{9.80 \times 10^{-5}}$ & $1.23 \times 10^{-4}$ & $3.49 \times 10^{-4}$ \\ 
   \hline
\end{tabular}
\end{center}
\end{footnotesize}
\end{table}

\begin{table}[htb]
\begin{footnotesize}
\begin{center}
\caption{"Covid" dataset (Entire): MSE for different levels of smoothing}
\label{covid-all-smoothing-table}
\begin{tabular}{lrrr}
  \hline
smooth & Only Fashion RP & Complete RP & Standard P\'olya \\ 
  \hline
no smooth & $2.46 \times 10^{-1}$ & $2.46 \times 10^{-1}$ & $2.50 \times 10^{-1}$ \\ 
  k = 3 & $\mathbf{3.98 \times 10^{-8}}$ & $7.37 \times 10^{-6}$ & $2.58 \times 10^{-3}$ \\ 
  k = 5 & $\mathbf{5.51 \times 10^{-8}}$ & $7.53 \times 10^{-6}$ & $2.64 \times 10^{-3}$ \\ 
  k = 10 & $\mathbf{1.54 \times 10^{-7}}$ & $8.63 \times 10^{-6}$ & $2.92 \times 10^{-3}$ \\ 
  k = 20 & $\mathbf{7.93 \times 10^{-7}}$ & $9.37 \times 10^{-6}$ & $3.10 \times 10^{-3}$ \\ 
  k = 30 & $\mathbf{1.06 \times 10^{-6}}$ & $9.80 \times 10^{-6}$ & $3.24 \times 10^{-3}$ \\ 
  k = 50 & $\mathbf{2.06 \times 10^{-6}}$ & $1.10 \times 10^{-5}$ & $3.46 \times 10^{-3}$ \\ 
   \hline
\end{tabular}
\end{center}
\end{footnotesize}
\end{table}

\begin{table}[htb]
\begin{footnotesize}
\begin{center}
\caption{"Covid" dataset (only Bots's posts): MSE for different levels of smoothing)}
\label{covid-onlyBOT-smoothing-table}
\begin{tabular}{lrrr}
  \hline
smooth & Only Fashion RP & Complete RP & Standard P\'olya \\ 
  \hline
no smooth & $2.45 \times 10^{-1}$ & $2.45 \times 10^{-1}$ & $2.49 \times 10^{-1}$ \\ 
  k = 3 & $\mathbf{3.23 \times 10^{-6}}$ & $5.33 \times 10^{-5}$ & $3.38 \times 10^{-3}$ \\ 
  k = 5 & $\mathbf{1.16 \times 10^{-5}}$ & $5.16 \times 10^{-5}$ & $3.38 \times 10^{-3}$ \\ 
  k = 10 & $\mathbf{2.84 \times 10^{-5}}$ & $6.88 \times 10^{-5}$ & $3.53 \times 10^{-3}$ \\ 
  k = 20 & $\mathbf{5.70 \times 10^{-5}}$ & $9.78 \times 10^{-5}$ & $3.80 \times 10^{-3}$ \\ 
  k = 30 & $\mathbf{1.67 \times 10^{-4}}$ & $1.81 \times 10^{-4}$ & $4.01 \times 10^{-3}$ \\ 
  k = 50 & $3.05 \times 10^{-4}$ & $\mathbf{2.94 \times 10^{-4}}$ & $4.38 \times 10^{-3}$ \\ 
   \hline
\end{tabular}
\end{center}
\end{footnotesize}
\end{table}

\section{Discussion and Conclusions}\label{conclusions}
Online Social Networks (OSN) represent a perfect environment for the study of the emotional reaction to public events. 
It has been observed that the sentiment of a message may be a driver for the diffusion of a message in online social networks~\cite{Zollo2015,Qiu2017, DelVicario2016c, Zollo2018a}. Interestingly, Ref.~\cite{Zollo2015} shows that, on different arguments, the \emph{sensitivity}, i.e. the emotional reaction to the event, finds a sort of stability.

Leveraging on this feature of the online debate, we apply here a modification of the P\'olya urn model, embedding a ``local" reinforcement effect~\cite{ale-cri-RP, ale-cri-GRP}, representing a sort of ``fashion" contribution and capturing the persistence of a common sentiment.  
%
%
Similarly to the standard  P\'olya urn, the future outcome depends on the entire story, but, differently from the original model, in the Rescaled P\'olya urn, the influence of the recent outcomes has a greater impact on future extractions. This represents the ``fashion" effect and its introduction   
%
%
properly captures the evolution of the sentiment of the online debate. The results collected in Section 
\ref{applications} show that indeed the Rescaled P\'olya model outperforms greatly the standard P\'olya model. Moreover, the RP urn model permits to have reliable predictions from past observations: in particular, the model parameters are fitted using the data from all the previous slots, thus capturing a sensitivity to a certain emotional reaction for the argument, as in~\cite{Zollo2015}. In this sense, when we focus on a single argument, as in the case of the Migration or Covid data sets, we have better results than when the argument is not specified, as in the case of 10 days traffic data set.


\indent Summarising, the present paper has essentially two targets: to propose a novel model for the prediction of the sentiment in the online debate and to examine and study the implications of the Rescaled P\'olya urn. Building a simple and realistic model improves our understanding of the phenomenon: in the particular case, the presence of a local reinforcement, i.e. the ``fashion" effect described above, shows how persistent is the emotional reaction to a public event.
It is worth to be mentioned that the application to Online Social Media is one of the possible applications of the proposed RP urn model: due to its abstractness and generality, it can be applied to any kind of phenomenon showing a local ``fashion" behaviour. \textcolor{black}{As it can be observed from the evolution of the model  parameters, 
all of them converge smoothly to a fixed value. Estimating the parameters using only the closest slots, is going to be the target of near future research.}



\subsection*{\bf Declaration}

Giacomo Aletti and Irene Crimaldi contributed to the theoretical definition of the Rescaled P\'olya urn model; Fabio Saracco performed the data collection and cleaning; Giacomo Aletti performed the simulation; all the authors wrote, revised and approved the manuscript.
\\
\color{black}
\subsection*{Acknowledgments}

\noindent Giacomo Aletti is a member of the Italian Group ``Gruppo
Nazionale per il Calcolo Scientifico'' of the Italian Institute
``Istituto Nazionale di Alta Matematica'' and Irene Crimaldi is a
member of the Italian Group ``Gruppo Nazionale per l'Analisi
Matematica, la Pro\-ba\-bi\-li\-t\`a e le loro Applicazioni'' of the
Italian Institute ``Istituto Nazionale di Alta Ma\-te\-ma\-ti\-ca''. \textcolor{black}{The authors acknowledge Fabio Del Vigna, Marinella Petrocchi and Manuel Pratelli for the bot detection on the above mentioned data sets.}
\\

\subsection*{Funding Sources}

\noindent Irene Crimaldi and Fabio Saracco are supported by the Italian
``Programma di Attivit\`a Integrata'' (PAI), project ``TOol for
Fighting FakEs'' (TOFFE) funded by IMT School for Advanced Studies
Lucca.  Fabio Saracco acknowledges also support from the European Project SoBigData++ GA. 871042\\



\appendix

\section{Parameters evolution and choice}\label{appendix-parameters}
\begin{figure}[htb]
\begin{center}
\includegraphics[width=\textwidth]{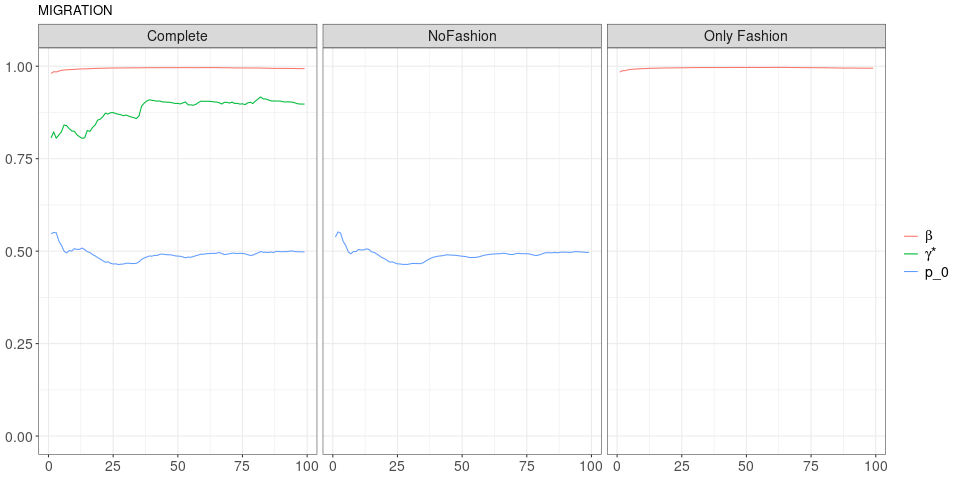}
\end{center}
\caption{``Migration" dataset (Entire): Model parameters evolution. All parameters are slowly varying and converging to stable values.}
\label{mi_de-all-parameters}
\end{figure}

\begin{figure}[htb]
\begin{center}
\includegraphics[width=\textwidth]{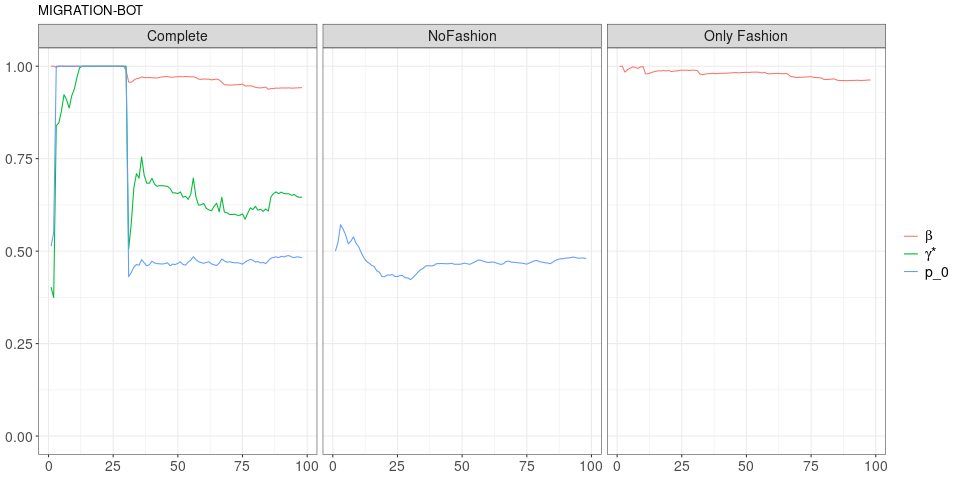}
\end{center}
\caption{``Migration" (only BOTs' posts): Model parameters evolution. In the left panel, it is possible to distinguish an ``Only fashion" initial phase ($\gamma^*\simeq 1$) and more balanced phase ($\gamma^*\in[0.5, 0.75]$).}
\label{mi_de-bot-parameters}
\end{figure}

As it is  mentioned in the main text, in order to fit the parameters of the model, we divided the entire dataset in time slots. Next, we use all slots previous to the one under consideration to fit the parameters. In this sense, we observe an evolution of the parameters as a matter of the evolution of the datasets, which is different when focusing on the different nature of users in the debate. Such a difference is particularly evident in the Migration debate. Human accounts show a nearly constant parameter dynamics: while $\beta$ is nearly constant in the Complete model, $\gamma^*$ and $p_0$ display a smooth slow variation of nearly the $10\%$ of their value, see Fig.~\ref{mi_de-all-parameters}. The dynamics of the parameters for automated accounts is completely different, see Fig.~\ref{mi_de-bot-parameters}: in the Complete model, while $\beta$ is slowly decreasing (but still experiencing a much greater decrease than the one observed for human accounts), parameters $\gamma^*$ and $p_0$  display a step-like dynamics, ending shortly after $S=25$.

\begin{figure}[htb]
\begin{center}
\includegraphics[width=\textwidth]{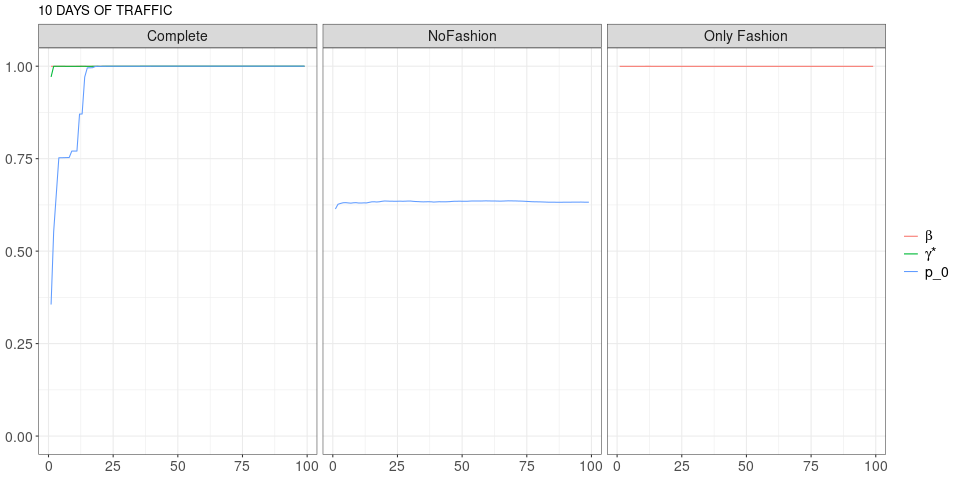}
\end{center}
\caption{``10 days of traffic" dataset (Entire): Model parameters evolution. All parameters for the complete model converge to 1 quite soon. With $\gamma^*\simeq 1$, the Complete model is practically equivalent to the ``Only Fashion" one.}
\label{10-all-parameters}
\end{figure}

\begin{figure}[htb]
\begin{center}
\includegraphics[width=\textwidth]{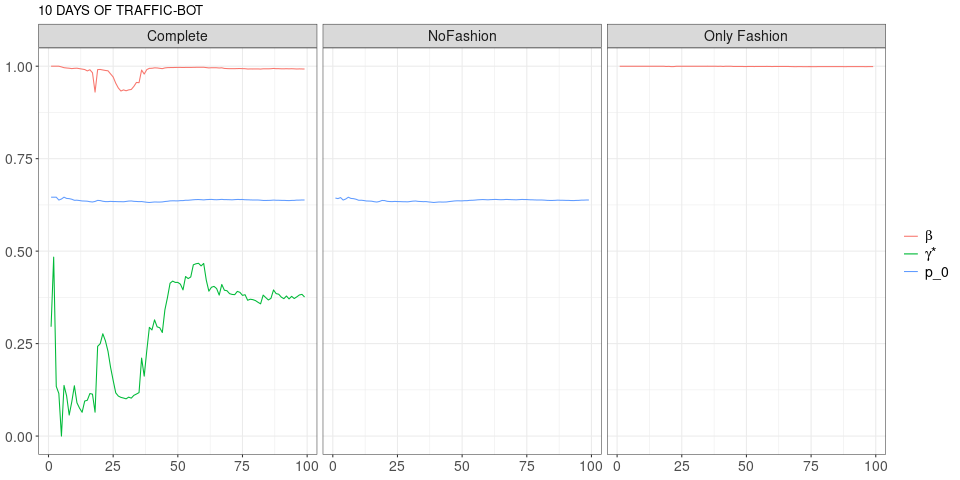}
\end{center}
\caption{``10 days of traffic" (only BOTs' posts): Model parameters evolution. Every parameter of the complete model, but $\gamma^*$, are essentially constant. Let us remark that $\gamma^*$ tunes the weight of the fashion process in the predictive mean.}
\label{10-bot-parameters}
\end{figure}

 A similar, but less evident, dynamics can be observed in the ``10 days of traffic" dataset, see Fig.~\ref{10-all-parameters}
 and \ref{10-bot-parameters}: in this case, all parameters converge to 1 quite soon in the case of the entire dataset. Instead, we can see that $p_0$ converge, but to something more than 0.6 quite immediately for the social bot subset, while the value of $\gamma^*$ oscillates between 0.5 and 0, before converging to something less than 0.4. $\beta$ is nearly 1 for both cases.

\begin{figure}[htb]
\begin{center}
\includegraphics[width=\textwidth]{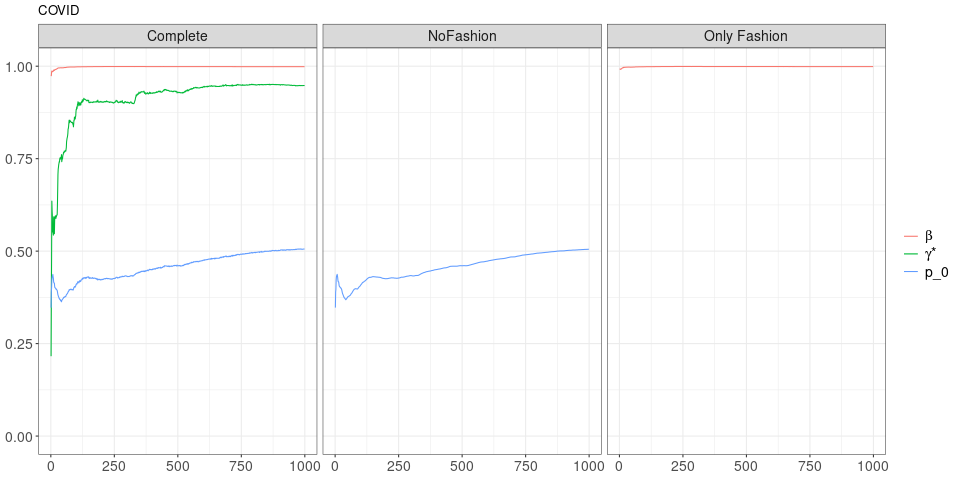}
\end{center}
\caption{``Covid" dataset (Entire): Model parameters evolution. All parameters are nearly constant or slowly converging. This slow convergence is probably due to the fact that the fit of the parameters is mostly driven by the long range history. In the case of the ``COVID" dataset, $S=1000$.}
\label{COVID-all-parameters}
\end{figure}

\begin{figure}[htb]
\begin{center}
\includegraphics[width=\textwidth]{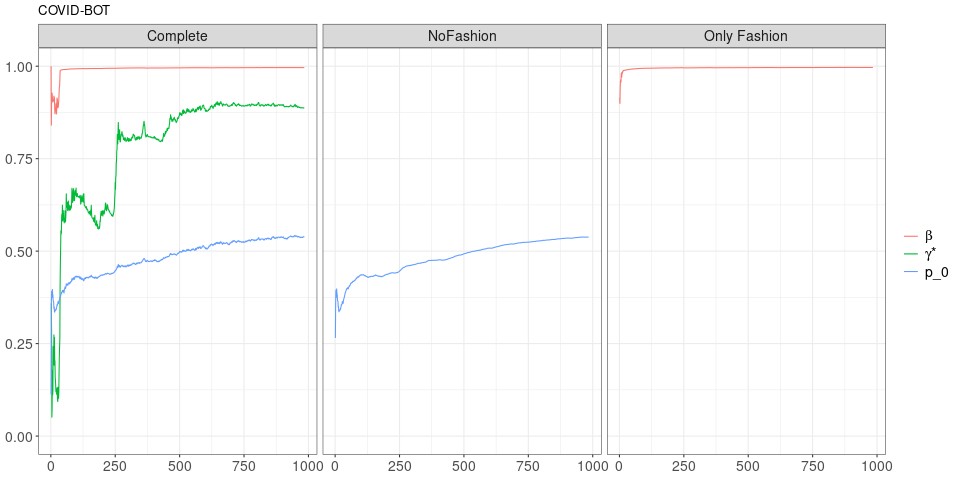}
\end{center}
\caption{``Covid" dataset (Only BOTs' posts): Model parameters evolution. Differently to the other datasets, the parameters evolution for the Bots' posts follows the one for the entire dataset, displaying a greater noise contribution for $\gamma^*$. In the case of the ``Covid" dataset, $S=1000$.}
\label{COVID-bot-parameters}
\end{figure}

In the case of the online debate during the COVID-19 epidemic, Fig.s~\ref{COVID-all-parameters} and \ref{COVID-bot-parameters}, the differences are extremely low, with the values of $\gamma^*$ quite flickering before converging to a value little lower than the one obtained for the entire dataset. All other parameters are quite similar, both in the value and in the dynamics.

\textcolor{black}{Actually, it is not clear if this evolution of the parameters for the BOT subset is due to the limited numerosity of the dataset or to an indeed different dynamics. This issue is not the focus of this work  and it is going to be the target of further analyses.}

\color{black}

\bibliographystyle{naturemag}

\end{document}